\documentclass[useAMS,usenatbib]{mn2e}
\usepackage{aas_macros,graphicx,times,multirow,amsmath,bbold}
\usepackage[]{color}
\usepackage[draft]{hyperref}

\title[Absorption effects on BH accretion disc emission]{Spectral and polarization properties of black hole accretion disc emission: including absorption effects}
\author[R. Taverna et al.]{R. Taverna$^{1}$\thanks{E-mail:
    \href{mailto:taverna@fis.uniroma3.it}{taverna@fis.uniroma3.it}},
  L. Marra$^{1}$, S. Bianchi$^{1}$, M. Dov\v{c}iak$^{2}$, R. Goosmann$^{3}$, F. Marin$^{3}$, G. Matt$^{1}$
  \newauthor and W. Zhang$^{2}$
\\ \\$^1$Dipartimento di Matematica e Fisica, Universit\`{a} degli Studi Roma Tre, via della Vasca Navale 84, I-00146 Roma, Italy\\
$^2$Astronomical Institute, Academy of Sciences of the Czech Republic, Bo\v{c}n\'{i} II 1401, CZ-14100 Prague, Czech Republic\\
$^3$Observatoire Astronomique de Strasbourg, Universit\'{e} de Strasbourg, CNRS, UMR, 7550, 11 rue de l'Universit\'{e}, F-67000 Strasbourg, France
}

\date{Accepted \ldots. Received \ldots; in
original form \ldots} \pagerange{\pageref{firstpage}--\pageref{lastpage}} \pubyear{2020}

\def\LaTeX{L\kern-.36em\raise.3ex\hbox{a}\kern-.15em
    T\kern-.1667em\lower.7ex\hbox{E}\kern-.125emX}

\def\maxrm {\mathrm{max}}
\def\minrm {\mathrm{min}}
\def\der {\mathrm{d}}

\def\rms {r_{\rm ms}}

\def\rg {r_{\rm g}}

\begin{document}

\label{firstpage}
\maketitle
\begin{abstract}
The study of radiation emitted from black hole accretion discs represents  
a crucial way to understand the main physical properties of these sources,
and in particular the black hole spin. Beside spectral analysis, polarimetry 
is becoming more and more important, motivated by the development of new 
techniques which will soon allow to perform measurements also in the X- 
and $\gamma$-rays. Photons emitted from black hole accretion discs in the 
soft state are indeed expected to be polarized, with an energy dependence 
which can provide an estimate of the black hole spin. Calculations performed 
so far, however, considered scattering as the only process to determine the 
polarization state of the emitted radiation, implicitly assuming that the 
temperatures involved are such that material in the disc is entirely ionized. 
In this work we generalize the problem by calculating the ionization structure 
of a surface layer of the disc with the public code {\sc cloudy}, and then 
by determining the polarization properties of the emerging radiation using the 
Monte Carlo code {\sc stokes}. This allows us to account for absorption effects 
alongside scattering ones. We show that including absorption can deeply modify 
the polarization properties of the emerging radiation with respect to what 
is obtained in the pure-scattering limit. As a general rule, we find that 
the polarization degree is larger when absorption is more important, which
occurs e.g. for low accretion rates and/or spins when the ionization of the 
matter in the innermost accretion disc regions is far from complete.

\end{abstract}

\section{Introduction}
\label{intro}

The black hole spin in accreting stellar-mass black hole binary systems is
currently estimated by using spectroscopic (either the iron $K\alpha$ line 
profile or the thermal disc continuum emission) or timing (kHz QPOs) techniques 
\cite[see][and references therein]{rey19}. A fourth technique, based on the
energy dependence of the polarization degree and angle of the thermal disc
emission, has been proposed in the late 70s \cite[][]{cs77,sc77,cps80}, and 
then revisited more recently \cite[][]{dov+08,lnm09,sk09,tav+20}. The renewed 
interest in the polarimetric technique is due to, on one hand, the discrepant
results provided by the other techniques in a few cases \cite[most notably 
GRO~J1655-40,][]{mot+14} and, on the other hand, to the re-opening of the X-ray 
polarimetric observing window provided by missions like {\it IXPE} \cite[][]{weiss+13}, 
due to be launched in 2021, and on a more distant future by {\it eXTP} \cite[][]{zhang+19}.

This technique is based on the influence of strong gravity on the polarization 
degree and angle of radiation. The polarization angle, in particular, is expected 
to rotate along the geodesics; the rotation is larger for more energetic photons, 
because they are mostly emitted closer to the black hole. When convolving over 
the entire disc emission, an energy dependence of the polarization angle results. 
This effect is larger the larger the spin of the black hole (because of the 
lower value of the Innermost Stable Circular Orbit, ISCO), which can therefore 
be determined \cite[see in particular][]{dov+08}.

\citet{sk09} included in their analysis the contribution of returning radiation, 
i.e. photons that, following null geodesics in the space-time around the BH, 
return to the disc before being reflected (in the hypothesis of $100\%$ albedo) 
towards the observer. They showed that spectra and polarization observables can 
be deeply modified with respect to those of direct radiation only due to the effects 
of reflection at the disc surface. \citet{tav+20} investigated the effects of 
considering different scattering optical depths on the intrinsic polarization 
and calculated how the spectra and polarization observables of both direct
and returning radiation are modified when a more realistic albedo profile 
is used to characterize the disc surface. 

In all the previous works, however, the polarization of the thermal disc is 
either calculated assuming a pure scattering slab of material \cite[][]{dov+08} 
or the Chandrasekhar's (\citeyear{chan60})  prescription, valid for a plane-parallel,
semi-infinite scattering atmosphere. To date, a self-consistent 
treatment for studying spectral and polarization properties of stellar-mass BH 
accretion disc emission accounting for absorption effects alongside scattering 
ones is still lacking, although the method described in \citet{tav+20} to include 
a more realistic albedo profile for the disc surface can be considered as a first 
attempt in this direction.

In the present work we still adopt the scenario of a thermal disc covered 
by a surface atmosphere, as in previous ones. Here, however, we move a further 
step forward by including in the polarized radiative transfer calculations absorption 
effects in the partially-ionized slab material.
To this aim, we first model the ionization structure of an optically-thick, surface 
layer of the disc using the photo-ionization code {\sc cloudy} \cite[][]{fer+17}. 
Secondly, we solve the radiative transfer for photons propagating within this 
surface layer, exploiting the Monte Carlo code {\sc stokes} \cite[][see also 
\citealt{gg07}]{mar18}. While our physical assumption of an ionized slab 
above a black body emitting disc is certainly simplistic, it allows us to start 
addressing the issue of the effect of absorption on the polarization properties 
of the emerging radiation. A self-consistent, simultaneous treatment of the disc 
structure and of polarized radiative transfer is beyond the scope of the present 
paper. Spectra and polarization properties are provided at the source, without 
considering the general relativistic corrections that affect the photon transport 
to the observer; the complete description of photon spectra and polarization 
as they would be measured at infinity will be handled in a future publication. 

The plan of the paper is as follows: in section \ref{section:themodel} we recall
the general concepts of our theoretical model, and describe in more details the
numerical implementation in section \ref{section:numericalimpl}. Results and
comparisons with previous works are introduced in section \ref{section:results}.
Finally, we summarize our findings and present our conclusions in section 
\ref{section:discussion}.

\section{The model}
\label{section:themodel}
\begin{figure*}
\begin{center}
\includegraphics[width=17.5cm]{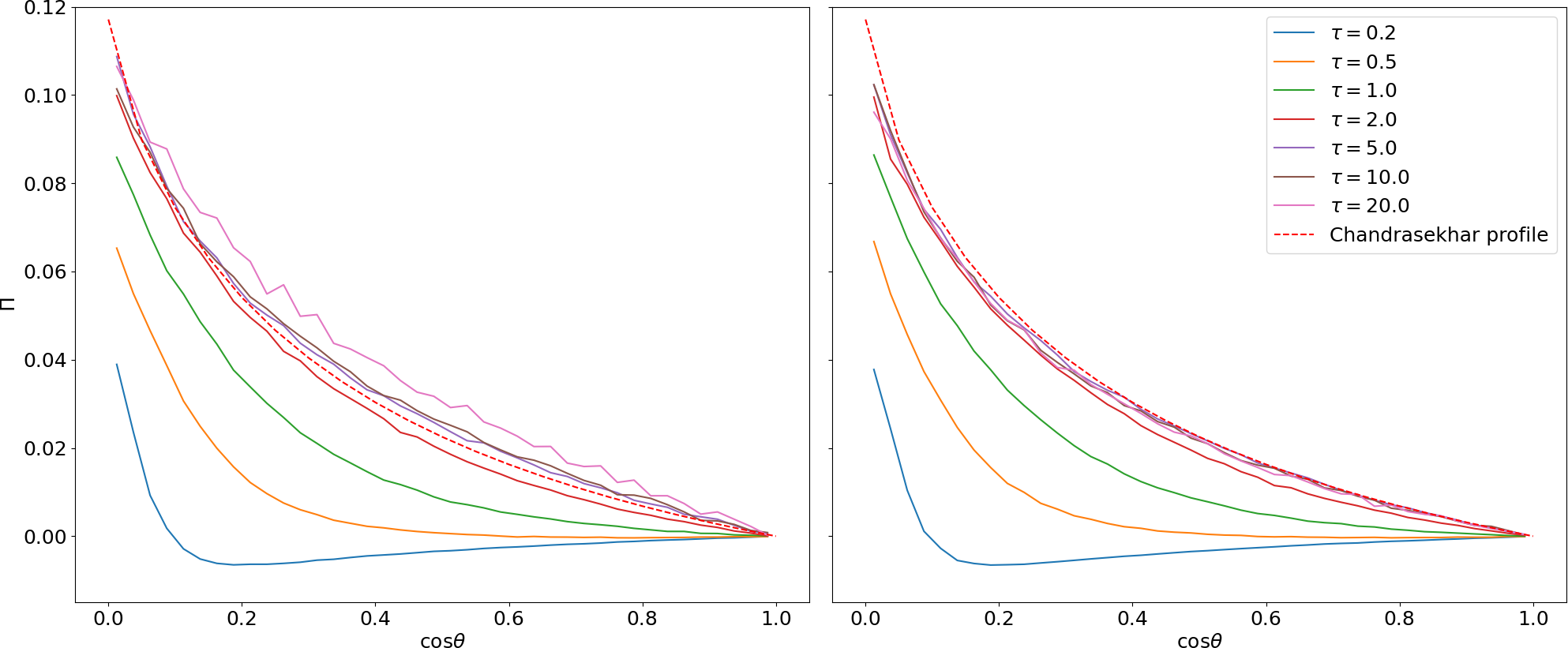}
\caption{Polarization degree in the pure-scattering limit, plotted as a function 
of the cosine of the inclination angle $\theta$ for isotropic emission from the
bottom of the disc surface slab. Positive values correspond to polarization parallel 
to the disc plane ($\chi=0^\circ$), while negative ones to polarization in the 
direction of the disc axis ($\chi=90^\circ$). Different values of the Thomson 
optical depth $\tau$ are explored: $0.2$ (cyan), $0.5$ (orange), $1$ (green), $2$ (magenta), 
$5$ (purple), $10$ (brown) and $20$ (pink). The polarization degree angular profile 
as described in \citet[see Table XXIV]{chan60} is also shown for comparison 
(red-dashed line). Stokes parameters have been summed over the $2$--$8$ keV 
(left-hand panel) and the $0.5$--$20$ keV (right-hand panel) energy bands.}
\label{figure:scattonly}
\end{center}
\end{figure*}
We recall in this section our basic assumptions. As described in previous works 
\cite[see e.g.][]{tav+20,dov+08}, we consider the accretion disc as a standard 
disc \cite[see][]{ss73}, with particles rotating around the center at the 
Keplerian velocity. The central, stellar mass BH is characterized by the values 
of its mass $M$ and dimensionless angular momentum $a$, with the space-time around 
described by the Kerr metric \cite[][]{nt73}. The disc stops at the radius of 
the ISCO $r_{\rm ms}$ \cite[see][]{bar+72}, and no-torque at the inner boundary 
is assumed. 

Internal viscous dissipations heat up the surface layers of the disc. We handle 
the distribution of the emitted photons through a local blackbody, with the
temperature $T$ varying with the radial distance $r$ from the center according 
to the Novikov-Thorne profile \cite[][]{nt73},
\begin{flalign} \label{eqn:NTtemperature}
T(\xi,M,\dot{M},a) &= 741f_{\rm col}\left(\frac{M}{\rm M_\odot}\right)^{-1/2}
\left(\frac{\dot{M}}{\rm M_\odot\,yr^{-1}}\right)^{1/4} \nonumber \\
\, & \,\,\,\,\,\,\times[f(\xi,a)]^{1/4}\,{\rm keV}\,; &
\end{flalign}
here $\xi=(r/\rg)^{1/2}$ (with $\rg=GM/c^2$ the gravitational radius), $\dot{M}$ 
is the BH accretion rate and $f$ is a function of $\xi$ and the BH spin $a$ 
\cite[see][for the complete expression; see also \citealt{pt74,wang00}]{tav+20}. 
In equation (\ref{eqn:NTtemperature}) we account for the energy shift of photons 
due to the scatterings they undergo with particles in the deepest layers of the 
disc through the hardening factor $f_{\rm col}$ \cite[][]{st95,dov+08,dea19}. 

The density of the disc material is modelled according to the radial profile
discussed by \citet{co17}, who provide the expressions for the total (hydrogen)
density $n_0(\rm H)$ at the equatorial plane \cite[see][for more details]{tav+20}. 
For ease of reading we reported the main formulae in Appendix \ref{appendix:densities}.
In order to obtain the corresponding values of the 
density at the disc surface, we assume for the sake of simplicity a Gaussian 
prescription for the vertical structure, so that
\begin{flalign} \label{eqn:nHsurface}
n({\rm H},z_*) &=n_0({\rm H})\exp\left(-\frac{z_*^2(r)}{h^2}\right)\,, &
\end{flalign}
where $h$ is the typical height of the disc at the radial distance $r$ and $z_*$ 
is the altitude above the disc equatorial plane at which the scattering optical 
depth calculated up to infinity is equal to 1 \cite[][]{tav+20}.

We solve the ionization structure of the disc only in its surface layer, in 
which the emitted blackbody radiation is processed. To describe the matter 
that composes this layer, we adopted the typical solar abundance 
\cite[][]{asp+05}, focussing in particular on the elements with $Z=1$ 
(hydrogen), $2$ (helium), $6$ (carbon), $7$ (nitrogen), $8$ (oxygen), $10$ 
(neon), $14$ (silicon), $16$ (sulfur) and $26$ (iron), neglecting the presence 
of dust. 

\section{Numerical implementation}
\label{section:numericalimpl}
In order to reproduce both the spectral and polarization properties of the
radiation emerging from the accretion disc, we use the Monte Carlo code {\sc 
stokes} \cite[][see also \citealt{mar+12,gg07}]{mar18}. This code was 
originally developed to solve the radiative transfer of near-IR to UV photons 
propagating inside regions of material where scattering represents the main 
source of opacity (like e.g. in AGN). However, with the purpose of adapting the 
code to our case, we resorted to the upgraded version 2.33 of {\sc stokes}, 
optimized for modeling the X-ray radiation propagating through a stratified, 
plane-parallel atmosphere, which fully adapts to describe the surface layer 
of the disc according to our model. In order to solve the atomic structure 
in this atmospheric layer, we use the version 17.01 of {\sc cloudy} \cite[last 
described in][]{fer+17}, an open-source, photo-ionization code to simulate 
the relevant processes that occur in astrophysical clouds. In particular, 
since we restrict our investigation to the case in which collisions are the 
dominant process for ionizing matter, we used the {\tt coronal} model, 
which allows us to calculate the ionization properties of a material slab by 
specifying only the (kinetic) temperature and the total (hydrogen) density of 
the gas. 

The surface atmospheric layer is, then, divided into a number $N_{\rm r}$ of 
patches, each one at a different radial distance $r$ from the central BH. For 
the sake of simplicity, we associate to each radial patch the corresponding 
values of temperature and density at the disc surface, as provided by equations 
(\ref{eqn:NTtemperature}) and (\ref{eqn:n0Hinner})--(\ref{eqn:n0Houter}), 
respectively, and we consider them as constant within the same patch. After 
having defined an $xyz$ reference frame, with the $z$-axis chosen along the 
disc symmetry axis, the code solves the ionization structure inside the patch 
by dividing it into several slices, characterized by the height $z$ with respect 
to the base of the surface layer. The maximum height $z_{\rm max}$ that the 
code reaches in each run is univocally determined by the input parameter 
$N_{\rm H}^{\rm stop}$, i.e. the hydrogen column density at which calculations 
are stopped. For every radial distance $r$ and altitude $z$, we finally extract, 
from the code output, the fractional abundance of each element in different 
states of ionization,
\begin{flalign} \label{eqn:fractionalabundance}
F^{\rm (i)}_{\rm X^{\alpha}}&=\frac{n^{\rm (i)}_{\rm X^\alpha}}{n^{\rm tot}_
{\rm X}}\,. &
\end{flalign}
In equation (\ref{eqn:fractionalabundance}), $n^{\rm (i)}_{\rm X^\alpha}$ is the 
number density of the ionic species $X^\alpha(i)$, with $\alpha$ the ionic charge
and $i$ the excitation level of the outermost electron, while $n^{\rm tot}_{\rm X}$ 
denotes the total number density of the element $X$.

The output of {\sc cloudy} obtained for each single radial patch is hence 
processed by a specific {\sc c++} script, in order to produce a suitable input 
file for {\sc stokes}, containing all the relevant information about the 
ionization structure. The emission region, from which photons are injected 
inside the layer can be featured as well in this input file, choosing between 
different geometries of emission. For our simulations we assumed the emitting 
source to be located at the bottom of the atmospheric layer ($z=0$); no bulk 
motion is considered for the emission 
region. Following our model prescriptions (see \S \ref{section:themodel}), for 
each radial patch of the surface layer we imposed that all the seed photons are 
emitted according to an isotropic blackbody at the temperature $T(r)$ (see equation 
\ref{eqn:NTtemperature}). This is tantamount to say that the propagation directions, 
along which photons are launched, are sampled by the polar angles
\begin{flalign} \label{eqn:seedthetaphi}
\theta_{\rm e}&=\arccos(\sqrt{r_1}) \nonumber & \\
\phi_{\rm e}&=2\pi r_2 &
\end{flalign}
with respect to the $z$-axis and the $xz$-plane, respectively; in equation
(\ref{eqn:seedthetaphi}) $r_1$ and $r_2$ are uniform deviates between $0$ 
and $1$. Seed radiation is set as unpolarized, i.e. the photon Stokes vectors 
are initialized to
\begin{flalign} \label{eqn:initstokesvector}
\left(\begin{array}{c}
i_0 \\ q_0 \\ u_0 \\ v_0
\end{array}\right)&=\left(\begin{array}{c}
1 \\ 0 \\ 0 \\ 0
\end{array}\right)\,. &
\end{flalign}

\begin{figure*}
\begin{center}
\includegraphics[width=17.5cm]{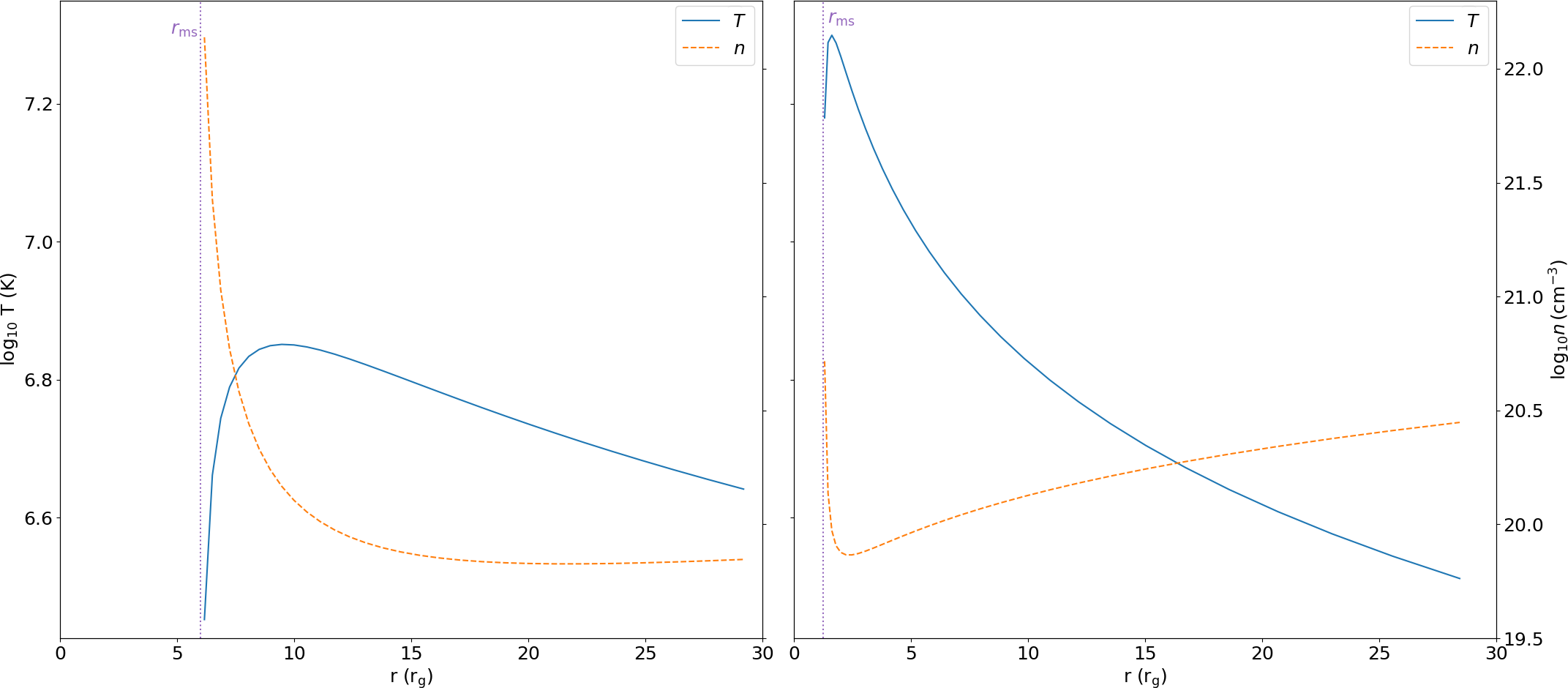}
\caption{Surface temperature (blue, solid line) and density (orange, dashed 
line) radial profiles obtained according to equations (\ref{eqn:NTtemperature}) 
and (\ref{eqn:n0Hinner})--(\ref{eqn:n0Houter}), respectively, in the case of a 
non-rotating (left-hand panel) and a maximally-rotating (right-hand panel) BH 
with mass $M=10\,{\rm M}_\odot$ and accretion rate $\dot{M}=0.1\,\dot{M}_{\rm 
Edd}$. The hardening factor $f_{\rm col}$ is set to $1.8$.}
\label{figure:tempdens1}
\end{center}
\end{figure*}

Photons are, then, followed along their trajectory, accounting for all the 
possible interactions (such as multiple scattering, free-free interactions 
or photoelectric absorptions) they can experience in the surface layer. All 
the photons which are not absorbed inside the layer are eventually collected 
in different virtual detectors, each one identified by the inclination $\theta$ 
and the azimuth $\phi$ which characterize the corresponding viewing direction 
in the $xyz$ frame. The total number of virtual detectors is fixed by specifying 
in input the numbers of points $N_\theta$ and $N_\phi$ of the $(\theta,\phi)$ 
angular mesh. For each detector, the Stokes parameters of the photons collected 
along the corresponding viewing direction are summed together, after having 
rotated the different Stokes parameter reference frames around the detector 
line-of-sight, to match with the detector frame. The final output of each run 
(corresponding to each radial distance $r$) consists of the Stokes parameters 
$i$, $q$, $u$ and $v$ of the emerging radiation as functions of the photon 
energy $E$ and of the two viewing angles $\theta$ and $\phi$. Given the axial 
symmetry of the adopted geometry, we actually integrated over $\phi$. The 
resolution $N_{\rm E}$ and the boundaries $E_\minrm$--$E_\maxrm$ of the photon 
energy band, as well as the number $N_{\rm phot}$ of seed photons to be launched 
in each single run, can be defined at the beginning of the {\sc stokes} input 
file. The linear polarization fraction $\Pi$ and the polarization angle $\chi$ 
are eventually obtained through the usual expressions
\begin{flalign} \label{eqn:polarizationobservables}
\Pi&=\frac{\sqrt{q^2+u^2}}{i} \nonumber & \\
\chi&=\frac{1}{2}\arctan\left(\frac{u}{q}\right)\,, &
\end{flalign}
where $\chi=0$ corresponds to polarization vectors oriented as perpendicular
to the $z$ axis (i.e. lying in the plane of the disc), increasing in the
clockwise direction.

\begin{figure*}
\begin{center}
\includegraphics[width=17.5cm]{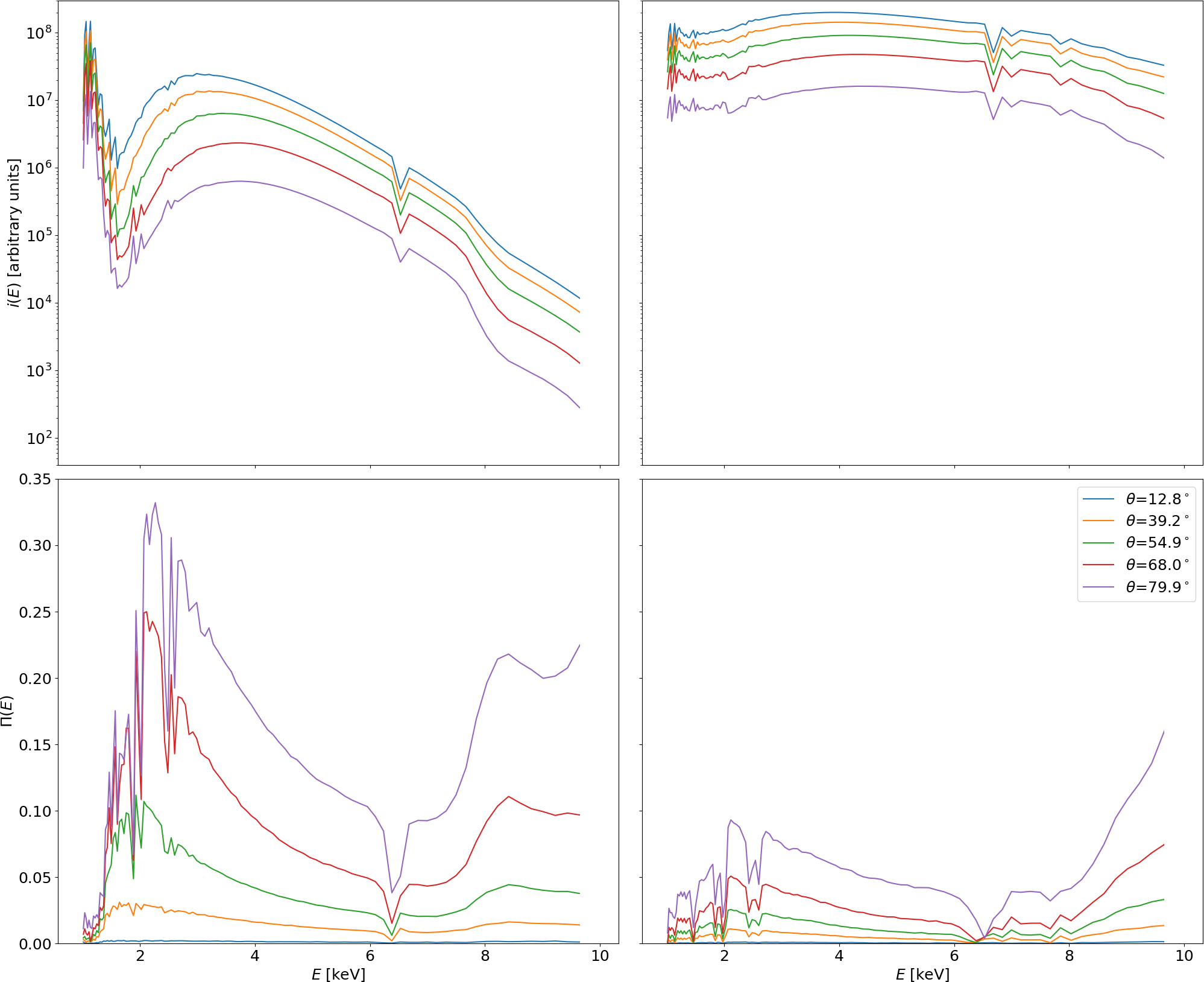}
\caption{Emerging spectra (top row) and polarization degree (bottom row) plotted 
for the two cases $a=0$ (left-hand column) and $a=0.998$ (right-hand column) and 
for $\theta=12.8^\circ$ (cyan), $39.2^\circ$ (orange), $54.9^\circ$ (green),
$68.0^\circ$ (red) and $79.9^\circ$ (purple). Here $M=10\,{\rm M}_\odot$, $L=0.1
\,L_{\rm Edd}$, $N_{\rm H}^{\rm stop}=10^{24}$ cm$^{-2}$ and $f_{\rm col}=1.8$.
Stokes parameters, weighted for the area and the temperature of each radial patch, 
are summed over the radial distance $r$ (see equations 
\ref{eqn:weightedstokes}).}
\label{figure:iPE1}
\end{center}
\end{figure*}

Finally, it should be noted that, by construction, at each radius the polarization 
is calculated by {\sc stokes} assuming a slab with constant density and temperature 
which is indefinite in the $xy$-plane. As we are dealing with a geometrically thin
accretion disc solution, we consider this approximation, which simplifies significantly
our computations, as acceptable.

\section{Results}
\label{section:results}
In this section we present the results of some significant simulations performed 
exploiting the codes illustrated in section \ref{section:numericalimpl}. For the 
sake of comparison with previous works, we start discussing the outputs obtained 
from {\sc stokes} in the pure-scattering limit and then we continue considering 
the effects of ionization in the disc surface layer. In the following,
the angular dependence is resolved in the colatitude $\theta$, while data are 
summed over the azimuthal angle ($N_\phi=1$).

\subsection{Pure-scattering limit} \label{subsec:res0}
The behavior of the polarization degree, after switching-off all interaction 
processes apart Compton scattering, is shown in Figure \ref{figure:scattonly} 
as a function of the cosine of the inclination angle $\theta$ (sampled over 
a $40$-point grid), for different values of the (Thomson) optical depth $\tau$. 
For each value of $\tau$, $N_{\rm phot}=10^9$ photons are injected in the 
layer according to a blackbody distribution at the temperature $T=1$ keV. Positive 
values refer to polarization in the plane of the disc ($\chi=0^\circ$), while 
negative ones to perpendicular polarization ($\chi=90^\circ$). In the left-hand
plot Stokes parameters have been integrated in the $2$--$8$ keV energy band, 
since this is the typical range of operations of the new-generation X-ray 
polarimeters like {\it IXPE}. In the right-hand one, instead, the $0.5$--$20$ 
keV band has been considered, as the best compromise between the choice of an
energy range in which all the photons are included and the need of a reasonable
computational time.

The curves are obtained following the prescriptions discussed in \S 
\ref{section:numericalimpl}, i.e. assuming that photons are emitted isotropically 
from a pointlike source placed at the bottom of the atmospheric layer, which is
a semi-infinite, plane-parallel slab. In almost all the cases explored, polarization 
vectors turn out to be oriented as parallel to the plane of the disc ($\Pi\ga0$), 
with the maximum polarization degree attained at high inclinations and monotonically 
decreasing down to $\sim 0$ at smaller ones. The only exception occurs at small 
optical depths ($\tau\la 0.5$), for which polarization may become perpendicular 
to the disc plane at small inclination angles. In particular, $\Pi$ assumes negative 
values for $\theta\la 53^\circ$ when $\tau=0.5$, while polarization vectors are 
definitely oriented in the direction of the projected disc symmetry axis for practically 
the entire range of inclinations for $\tau=0.2$. The effect can be explained as 
follows. When small values of $\tau$ are considered, photons which are more likely 
to be scattered are those emitted at large inclinations with respect to the slab 
normal, since they experience a larger effective optical depth. Because the electric 
vector of the scattered photons oscillates perpendicularly to the scattering plane, 
these photons will emerge with orthogonal polarization. On the other hand, photons 
originally emitted at small inclinations will be practically unpolarized if they 
emerge at small inclinations too, while, for symmetry reasons, their polarization 
is expected to be very large and oriented in the plane of the disc if they emerge 
at large inclinations. As a result, polarization turns out to be negative for a 
large interval of inclinations, becoming positive only close to $\cos\theta\sim0$. 
On the other hand, at large optical depths multiple scattterings can occur, which 
mitigate this behavior; moreover, scatterings become more and more frequent also 
for photons emitted at small inclinations, which contribute much more to the overall polarization pattern. The emergence of negative polarization has been originally 
noted by \citet{nag62} and \citet{gs78}, albeit in a slightly different scenario. 
In particular, the latter authors showed that, while only positive polarization 
is present if the photons are preferentially emitted deep in an accretion disc,
negative polarization arises when the source function is almost constant along 
the disc vertical structure. In the latter case, as \citet{gs78} explain, many 
photons before the last scattering travel almost parallel to the disc surface, 
and moreover their scattering angle is close to 90$^{\circ}$, a situation similar 
to ours in the case of small optical depths.

Increasing the layer optical depth, the angular behavior of the polarization 
fraction approaches, as expected, the classic solution obtained by \citet{chan60} 
assuming a semi-infinite slab (red-dashed line). In this condition, photons can 
emerge from the upper boundary of the layer only after a large number of scatterings, 
at variance with what happens for small optical depths, when a significant number 
of photons can escape even without suffering any scattering (and then remaining 
unpolarized). It is interesting to note that the polarization degree attained for 
optical depths larger than $\sim 5$ turns out to even slightly exceed that expected 
according to the Chandrasekhar's profile, as it can be seen looking at the left-hand
panel in Figure \ref{figure:scattonly}. The reason is that, while the Chandrasekhar's 
solution is obtained assuming elastic (Thomson) scattering, in our {\sc stokes} 
model Compton downscattering  is accounted for\footnote{We point out that Compton
up-scattering is not yet implemented in the current version of {\sc stokes}}. 
As a consequence of the energy 
shift toward lower energies, a certain number of photons drops out of the selected 
energy band, an effect of course increasing with the average number of scatterings 
(and then with $\tau$). Since the photons that are lost in this way are those 
which have suffered more interactions, and therefore are more isotropised and 
less polarized, an increase of $\Pi$ with respect to \citet{chan60} can be observed 
when the energy band is restricted to the $2$--$8$ keV range. Results closer 
to those given by the Chandrasekhar's solution can be obtained, in fact, by 
considering a wider energy range, as shown in the right-hand panel of Figure
\ref{figure:scattonly}, where Stokes parameters are integrated over the $0.5$--$20$
keV band.
\begin{figure*}
\begin{center}
\includegraphics[width=17.5cm]{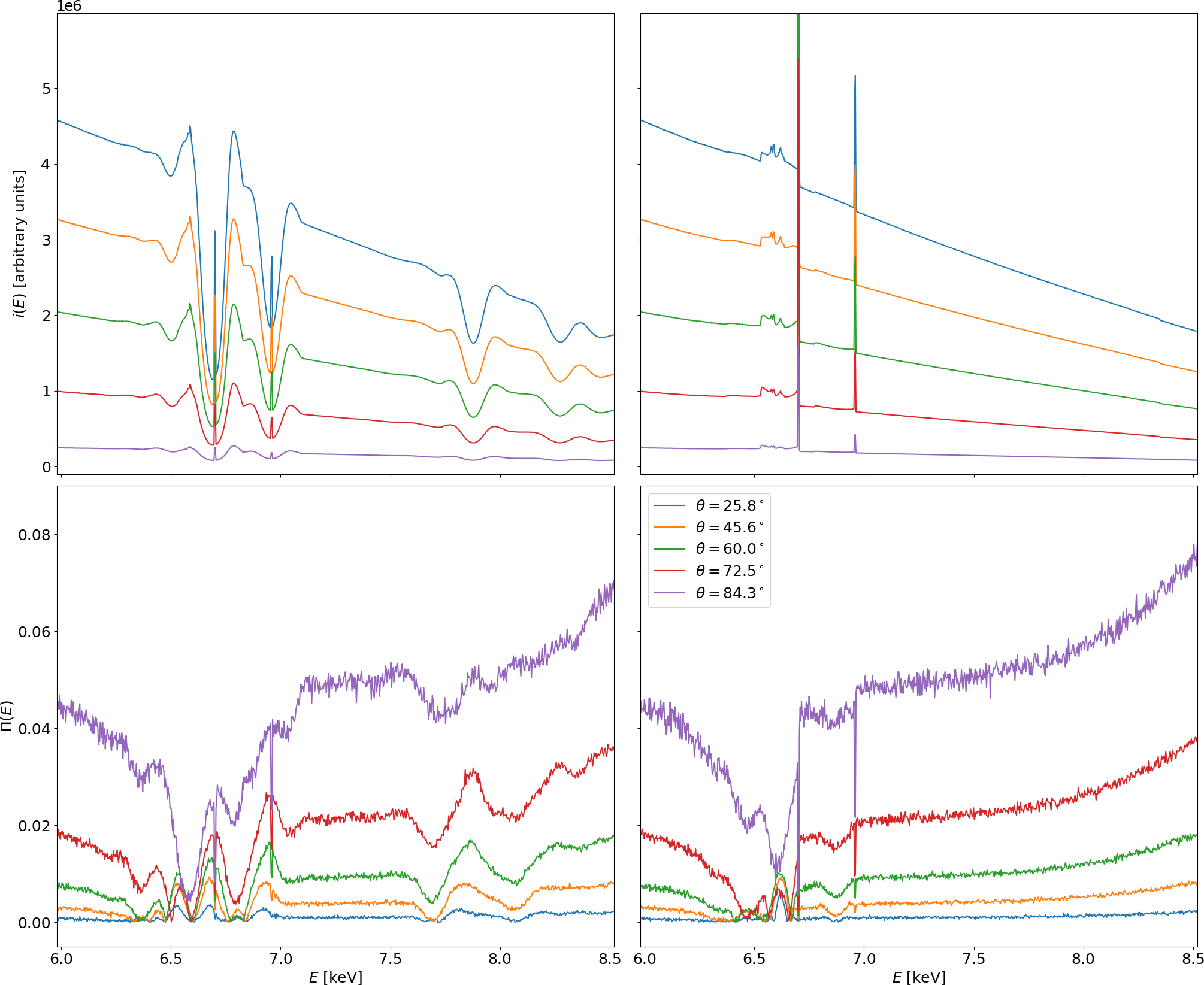}
\caption{Photon flux (top row) and polarization degree (bottom row) plotted as
functions of the photon energy (sampled with a 5000-point grid) in the case of 
$a=0.998$, for the first radial bin of the disc surface and for five different 
inclinations of the viewing direction. The values of the other parameters are 
the same as in Figure \ref{figure:iPE1}. Scattering, free-free, recombination
and photo-electric absorption effects are accounted for in the plots of the left
column, while resonant scattering lines are artificially turned off in those 
of the right one.}
\label{figure:features}
\end{center}
\end{figure*}
\begin{figure*}
\begin{center}
\includegraphics[width=17.5cm]{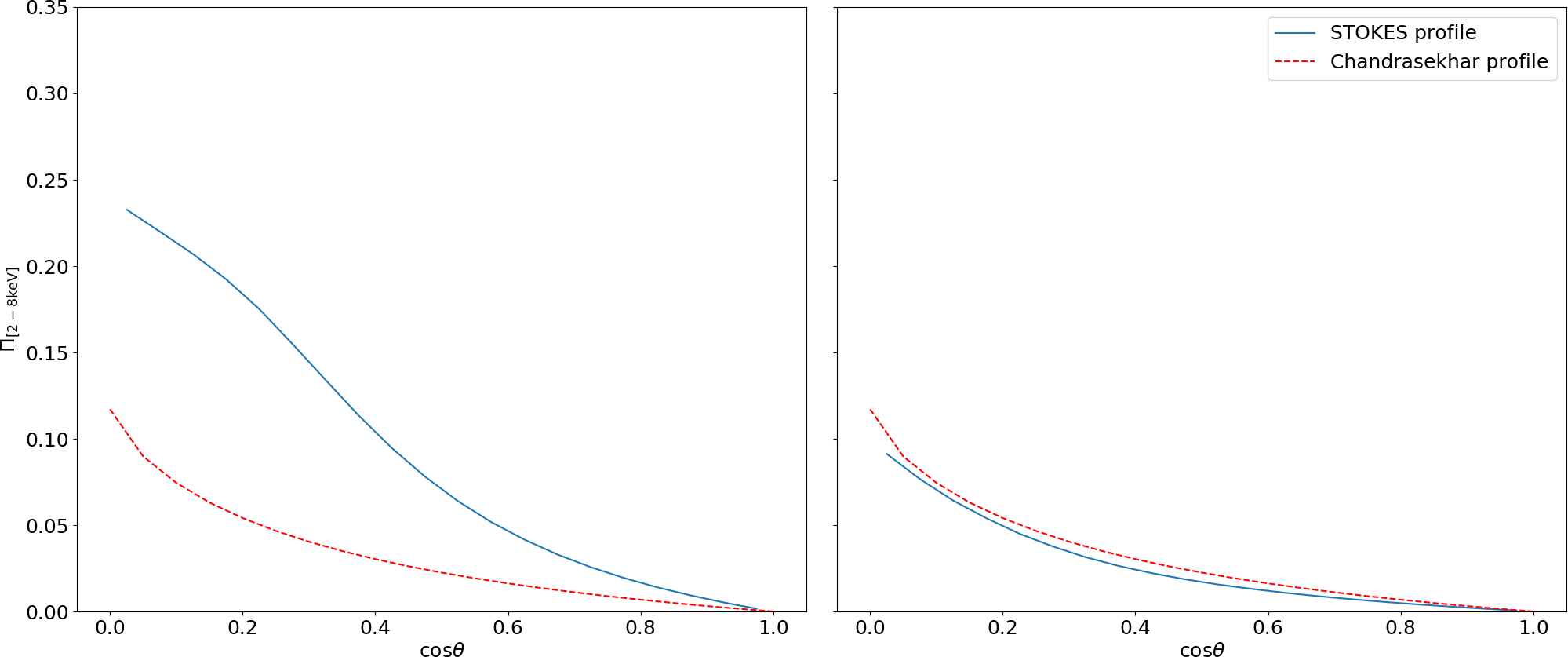}
\caption{Polarization degree plotted as a function of the cosine of the 
inclination angle $\theta$ for the two cases of $a=0$ (left-hand panel) 
and $a=0.998$ (right-hand panel) and the same values of parameters as in 
Figure \ref{figure:iPE1}. Stokes parameters have been summed over the radial 
distance from the central BH (see equations \ref{eqn:weightedstokes}) and 
energy integrated between $2$ and $8$ keV. Blue-solid lines mark the behavior 
predicted by {\sc stokes}, while the red-dashed lines that calculated according 
the formulae by \citet[see Table XXIV]{chan60}.}
\label{figure:Pchimu1}
\end{center}
\end{figure*}

A similar behavior to that just discussed has been already presented in \citet[see 
their Figure 1]{dov+08}, with some substantial differences: (i) the change of 
orientation of the polarization vectors occurred at quite larger optical depths
(for $\tau\sim2$); and (ii) higher polarization fractions were attained, especially 
for low optical depths, contrary of what happens in the present case. The main 
reason of these differences resides in the different layout adopted in previous 
works \cite[][see also \citealt{tav+20}]{dov+08} with respect to the present paper, 
with the emitting region located in the middle of the atmospheric layer (i.e. 
at $z=z_{\rm max}/2$ instead of $z=0$). In this situation, to ensure that photons 
are emitted isotropically in both the upper and the lower half-spaces of the layer, 
the distribution of the emission angles $\theta_{\rm e}$ should be corrected by
\begin{flalign} \label{eqn:seedthetaphi2}
\theta_{\rm e}&=\arccos(1-2r_1)\,, &
\end{flalign}
in place of that indicated in the first of equations (\ref{eqn:seedthetaphi}).
As a consequence, many more photons are originally emitted along the disc, rather
than perpendicularly to it, and therefore the polarization tends to be perpendicular
to the disc plane even for relatively large optical depths (for very large depths, 
however, the original distribution of photons is of course no longer important).

\subsection{Including the ionization structure} \label{subsec:res1}
We then included the ionization structure of the disc surface layer as calculated 
by {\sc cloudy}. We consider the two extreme cases of a non-rotating ($a=0$) and 
a maximally-rotating ($a=0.998$) BH, with mass $M=10\,{\rm M}_\odot$. We give the 
results for a portion of the disc between $r=\rms$ and $r=30\,\rg$, dividing the 
surface layer into $N_{\rm r}=30$, logarithmically-spaced radial bins. In this 
section we consider the set of parameters already used in previous works \cite[see 
e.g.][]{tav+20}, i.e. a hardening factor $f_{\rm col}=1.8$ and a mass accretion 
rate $\dot{M}$ chosen in such a way that the accretion luminosity\footnote{The 
calculation of the accretion efficiency is performed using the expressions reported 
in \citet{jp11}, see also \citet{kraw12}.} amounts to $10\%$ of the Eddington 
limit $L_{\rm Edd}$. The maximum height $z_\maxrm$ of the surface layer at each 
radial patch is chosen, instead, by setting the stop column density to $N_{\rm 
H}^{\rm stop}=10^{24}\,{\rm cm}^{-2}$, which corresponds to a Thomson optical 
depth $\tau\simeq 0.67$. The corresponding radial profiles of the surface temperature 
and density are plotted in Figure \ref{figure:tempdens1}. Also in this case, 
{\sc stokes} runs are performed launching $N_{\rm phot}=10^9$ seed photons for 
each radial patch, while Stokes parameters are sampled over the $1$--$10$ keV 
energy range through a $100$-point grid and over the $0$--$\pi/2$ inclination
interval through a 20-point grid, which a posteriori turned out to be a good 
compromise between the statistical significance and the computational time 
($\approx 1$ day to complete one run on an Intel i7, 4-core machine) for the 
chosen values of the parameters. 

\begin{figure*}
\begin{center}
\includegraphics[width=17.5cm]{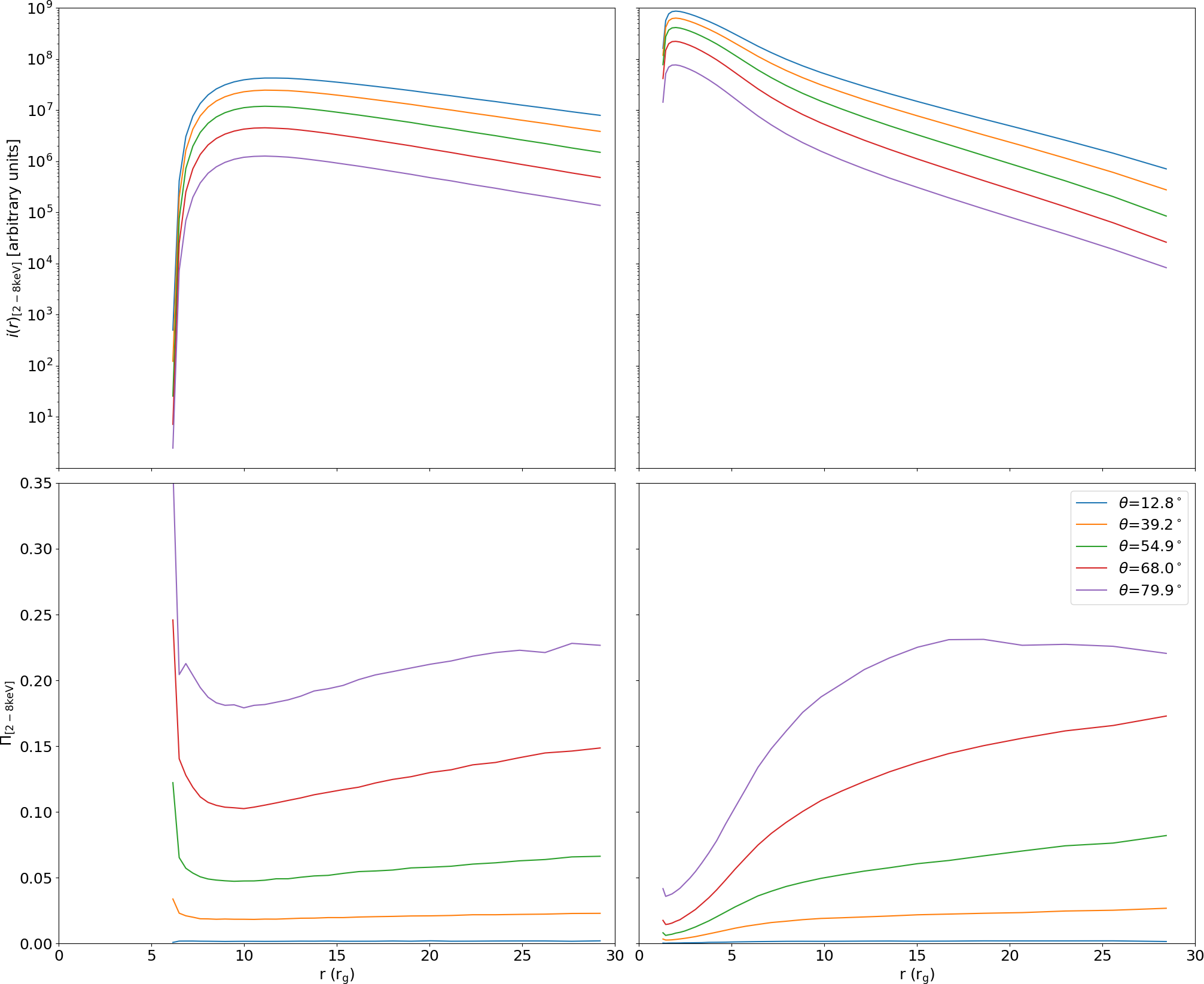}
\caption{Photon flux (top row) and polarization degree (bottom row) plotted as
functions of the radial distance $r$ in the two cases of $a=0$ (left-hand 
column) and $a=0.998$ (right-hand column), for the same values of parameters 
as in Figure \ref{figure:iPE1} and for five different inclinations of the 
viewing direction. Stokes parameters have been energy integrated between $2$ 
and $8$ keV.}
\label{figure:IPr1}
\end{center}
\end{figure*}

Figure \ref{figure:iPE1} shows the spectra and polarization degree obtained in
the two cases of $a=0$ and $a=0.998$, for five different viewing angles $\theta$. 
In order to correctly account for the different number of photons emitted from 
each radial bin, in these plots the Stokes parameter fluxes are summed 
over the radial distance $r$ in the following way,
\begin{flalign} \label{eqn:weightedstokes}
\bar{i}(E,\theta)&=\sum_r i(E,r,\theta)A(r)T^4(r) & \nonumber \\
\bar{q}(E,\theta)&=\sum_r q(E,r,\theta)A(r)T^4(r) & \nonumber \\
\bar{u}(E,\theta)&=\sum_r u(E,r,\theta)A(r)T^4(r)\,, &
\end{flalign}
where $A(r)=2\pi r\der r$ is the area of the annular radial patch, with width 
$\der r$, at the distance $r$ from the center, while the factor $T^4(r)$ takes 
into account the different surface temperatures which characterize each radial 
patch.

>From the top row of Figure \ref{figure:iPE1} one can clearly see that the
emerging photon flux is in general higher for a maximally rotating BH than for
the case of $a=0$. This can be easily explained by the fact that, for $a=0.998$, 
the disc extends up to regions much closer to the BH horizon; here temperatures 
are much higher than for the non-rotating case, peaking at $\sim 1.7$ keV 
against a maximum value $\sim 0.6$ keV for $a=0$ (see Figure \ref{figure:tempdens1}). 
As a consequence, for the Schwarzschild BH the seed photon blackbody peak can 
be expected to occur at $\sim 1$ keV, while starting from $3$--$4$ keV the strong 
energy dependence of the photoelectric absorption cross-section becomes more evident. 
On the other hand, the maximum of the injected blackbody falls at around $3$ keV 
in the maximally-rotating case, so that the contribution of seed photons is much 
more relevant all over the selected $1$--$10$ keV energy range. Absorption 
turns out to be important as well, as shown by the occurrence of several spectral 
features superimposed to the continuum in both the $a=0$ and $a=0.998$ cases. 
These lines appear to be mostly located at low energies ($1$--$2$ keV), with 
the exception of two, quite strong absorption features which appear at $\sim 
6.5$ keV and $\sim 8$ keV.

The correspondent polarization degrees are plotted as functions of the photon 
energy in the bottom row of Figure \ref{figure:iPE1}. Looking at the figures,
it is immediately clear that, as a general rule, polarization degree is higher
when the absorption is more relevant (see also the comparison with the pure 
scattering case). This pattern can be explained by noting that, when absorption 
is important, most of the emerging photons are those originally emitted almost 
vertically and which suffer only one scattering: those photons are all polarized 
with the polarization vector parallel to the disc surface. Photons originally
emitted at high inclinations are instead more likely to be absorbed; those photons, 
in the pure scattering limit, would provide mostly perpendicular polarization, 
with a reduction of the net polarization degree. Polarization of radiation 
emerging from the disc surface layer decreases by decreasing the observer's 
inclination angle, attaining a value close to $0$ at small $\theta$ ($\la 13
^\circ$) and at essentially all the photon energies. 
For $a=0$ (bottom-left panel), the polarization degree is in general higher than 
for the maximally-rotating BH case, except for the lower energies (at around $1$ 
keV), where $\Pi$ turns out to be very low (below $2\%$) at all the inclinations. 
This can be explained noting that, as mentioned above, primary photons peak indeed 
at such low energies, so that photons emerging at $1$--$2$ keV are essentially all 
seed photons, which are assumed to be unpolarized (see \S \ref{section:numericalimpl}). 
At higher energies the polarization fraction increases rapidly, up to a maximum 
value close to $\sim 30\%$ for the highest inclinations. This behavior corresponds 
to the most important decline of the photon flux at around $2$ keV. 
For $a=0.998$ (bottom-right panel), the energy dependent behavior of the polarization
degree closely follows that just discussed for the Schwarzschild case, notwithstanding
the lower values attained, which are in general reduced by a factor of $\sim 
2$--$3$ (with a maximum around $15\%$ at higher energies). As for the spectra, 
also the polarization fraction plots show further peculiar features between $6$ 
and $8$ keV, with the occurrence of quite narrow drops which seems to be associated, 
at first glance, to the analogous absorption lines in the flux. However, contrary 
of one could expect, the decrease in the photon flux would seem to correspond this 
time to a decrease also in the polarization degree.

\begin{figure*}
\begin{center}
\includegraphics[width=17.5cm]{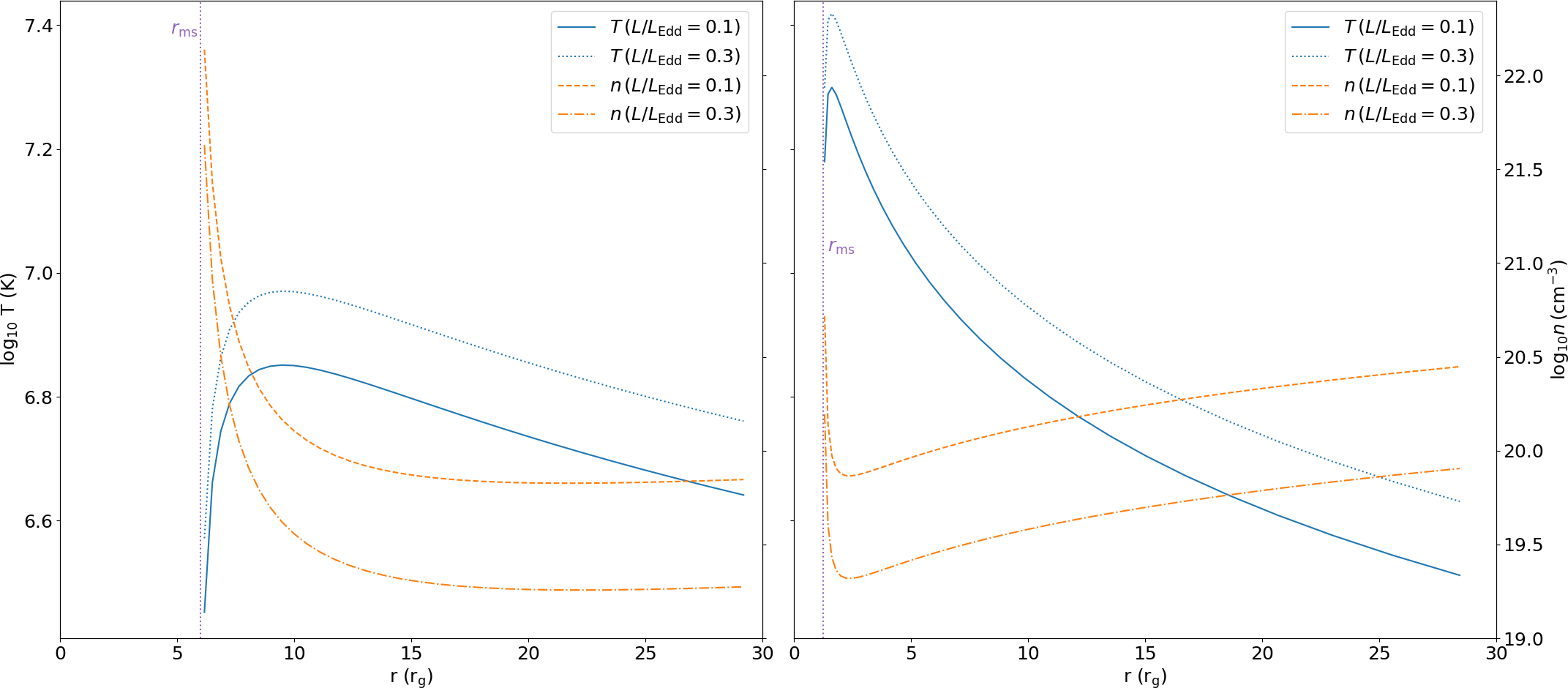}
\caption{A comparison of the temperature (blue) and density (orange) radial 
profiles in the two cases of $a=0$ (left) and $0.998$ (right), for $L=0.1\,
L_{\rm Edd}$ (the case also reported in Figure \ref{figure:tempdens1}) and 
$0.3\,L_{\rm Edd}$.}
\label{figure:tempdens2}
\end{center}
\end{figure*}

This counter-intuitive behavior can be explained by noting that both the spectral 
and the polarization profiles reported here are strongly affected by our choice
to adopt a 100-point energy grid between $1$ and $10$ keV. In order to better
understand the nature of these features we report, for the sake of example, the 
results of an additional run of {\sc stokes} in the case of $a=0.998$ and for
the first radial bin of the disc surface, sampling the energy range with a 5000-point
mesh. Figure \ref{figure:features} (left column) shows the plots of the spectrum 
and polarization degree for five different inclination angles and focussing on 
the energy band between $6$ and $8.5$ keV. Moreover, with the purpose of a greater 
clarity, we report in the right column the same situation as in the left one, 
in which we artificially removed all the features which are produced by resonant 
scattering. This allows one to clearly distinguish the contribution from the iron
line complex, which in the selected energy range is essentially due to the transitions 
of Fe$^{+24}$ and Fe$^{+25}$ in various ionization states \cite[][]{sar88,fab+00},
with two emission lines at $6.7$ and $7$ keV which stand out with respect to the
continuum.
Looking at the related polarization fraction plot (bottom-right panel), a sudden 
decrease of $\Pi$ occurs in correspondence with the aforementioned spectral features, 
as expected in the case of emission lines. Once resonant scattering is added in 
the computations (left panels), several absorption features appear in the spectrum, 
which are much broader than the narrow emission lines just discussed. In particular, 
two important absorption lines occur at exactly the same energies of the most 
important iron emission lines. A polarization degree growth correctly correponds 
to the scattering absorption lines, with the two significant drops at $6.7$ and 
$7$ keV due to the iron emission lines which are still visible in the middle 
of the polarization fraction peaks. We note that the features discussed here 
above can be fully resolved only 
if the energy grid is sufficiently fine (i.e. with more than $\sim 5000$ points 
between $1$ and $10$ keV). However, we remark that the energy resolution we adopted 
throughout the paper (i.e. 100 points in the $1$--$10$ keV range) is still far 
better than that of many existing instruments, and in particular of the forthcoming 
photoelectric polarimeters like {\it IXPE}. Furthermore, considering a too precise 
energy grid would lead to unacceptably long computational times, as well as to 
over-detailed spectral and polarization behaviors in the majority of the selected 
energy range, which are mostly pointless for our research. For these reasons we 
resort to the coarser, but still acceptable, 100-point energy resolution in the 
following.

\begin{figure*}
\begin{center}
\includegraphics[width=17.5cm]{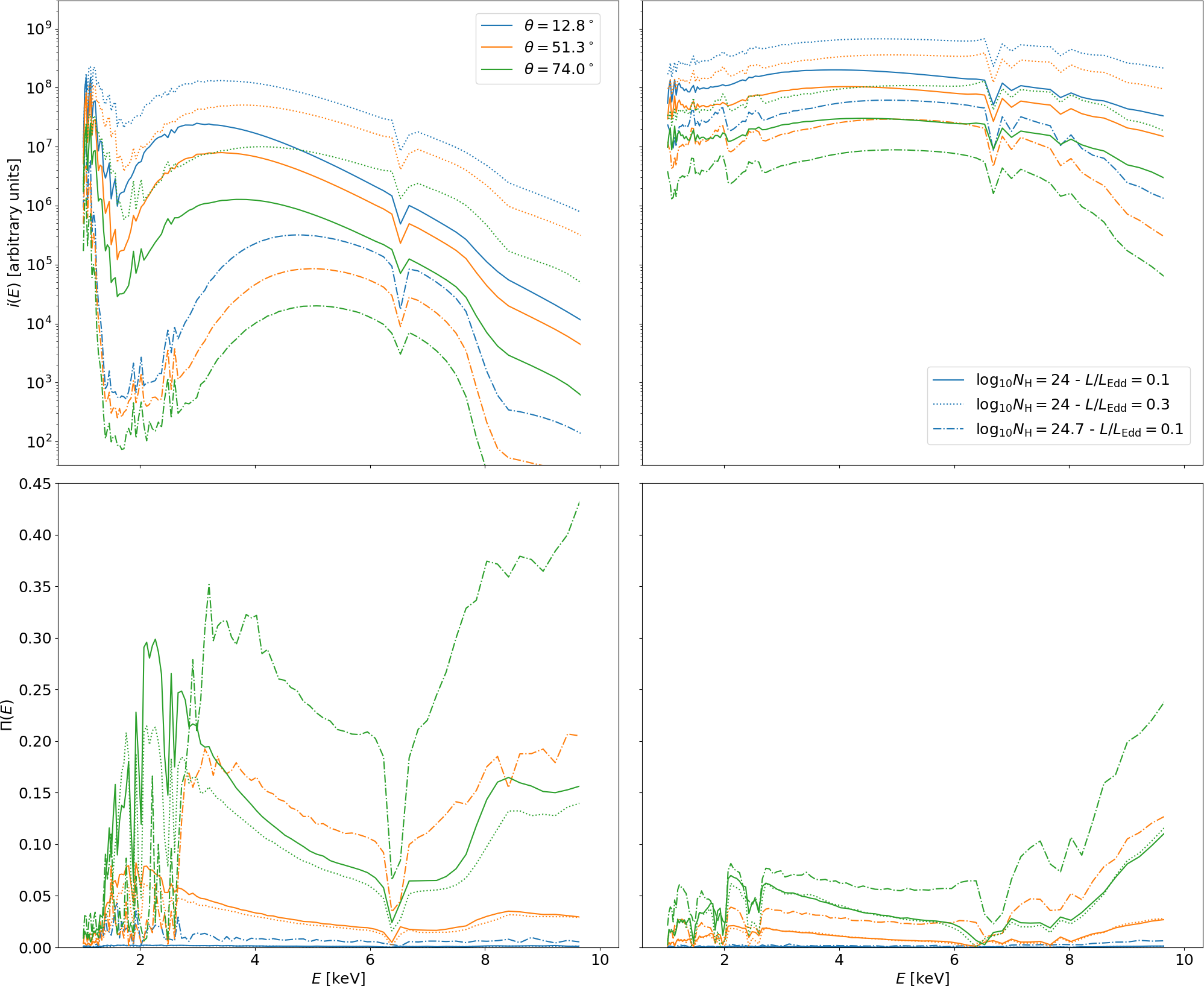}
\caption{Spectrum (top row) and polarization degree (bottom row) of the radiation 
emitted from the disc surface in the cases of $a=0$ (left) and $0.998$ (right), 
for $N_{\rm H}^{\rm stop}=10^{24}\,{\rm cm}^{-2}$ -- $L=0.1\,L_{\rm Edd}$ (solid 
lines), $N_{\rm H}^{\rm stop}=10^{24}\,{\rm cm}^{-2}$ -- $L=0.3\,L_{\rm Edd}$ 
(dotted lines) and $N_{\rm H}^{\rm stop}=5\times10^{24}\,{\rm cm}^{-2}$ -- $L=
0.1\,L_{\rm Edd}$ (dash-dotted lines), and for three different inclinations of 
the viewing direction: $\theta=12.8^\circ$ (blue), $51.3^\circ$ (orange) and 
$74.0^\circ$ (green). The values of the other parameters are chosen as in Figure 
\ref{figure:iPE1}.}
\label{figure:iPE2}
\end{center}
\end{figure*}

In order to better understand how the ionization structure of the disc surface 
layer determines the polarization pattern of the emerging radiation, Figure 
\ref{figure:Pchimu1} shows the behaviors of the polarization degree plotted
as a function of the cosine of the inclination angle $\theta$. Here the Stokes 
parameters, which are still summed over the radial distance as described above 
(see equations \ref{eqn:weightedstokes}), are further integrated over the photon 
energy in the $2$--$8$ keV {\it IXPE} energy band. The angular distribution 
of the polarization degree as predicted by Chandrasekhar's (\citeyear{chan60}) 
formulae is also reported in the plots (red-dashed lines). The fact that $\Pi$ 
assumes only positive values at all the inclinations considered (for both the 
cases of $a=0$ and $a=0.998$) shows that emerging radiation is mostly polarized 
perpendicularly to the disc symmetry axis. This, as already discussed in \S 
\ref{subsec:res0}, follows from the original assumption of radiation emitted 
isotropically from the base of the disc surface layer, so that photons are essentially 
emitted upwards, with propagation direction close to the disc axis. Furthermore, 
in the case of a maximally-rotating BH the polarization degree turns out to closely 
follow the Chandrasekhar's profile, contrary of what happens for $a=0$, where 
the polarization fraction largely exceeds, at low inclinations, that predicted 
by \citet{chan60}. Since the Chandrasekhar's profile is the reference one for 
models that assume scattering as the only process responsible for photon 
polarization\footnote{We note, however, that for the optical depth assumed 
in Figure \ref{figure:Pchimu1} (i.e. $\tau\simeq0.67$), the curve of the polarization 
degree in the pure-scattering limit stands below the Chandrasekhar's profile (see
e.g. Figure \ref{figure:scattonly}).}, one can safely conclude that the polarization 
properties in the $a=0.998$ case are mainly determined by scattering. On the other hand, 
absorption plays a more important role in the non-rotating case. This can be further 
confirmed looking again at the temperature and density distributions reported in 
Figure \ref{figure:tempdens1}. In fact, due to the higher temperature reached close 
to the BH horizon, the fraction of ionized atoms is clearly expected to be larger 
for $a=0.998$, while for $a=0$ absorption effects become more important. In this 
regard, one should also notice that the total density close to the ISCO is much 
larger (by a factor of $\sim 30$) for the non-rotating case than for the maximally 
rotating one; this enhances the effects of absorption for $a=0$. 

To complete this analysis, Figure \ref{figure:IPr1} shows the photon flux and
the polarization degree plotted as functions of the radial distance $r$ from 
the center, again for $a=0$ and $a=0.998$; also in this case Stokes parameters 
have been integrated over energy in the $2$--$8$ keV band. The behavior of the 
photon flux (top row) turns out to naturally follow the radial profile of the 
disc surface temperature reported in Figure \ref{figure:tempdens1}, peaking at 
the distance $r$ characterized by the maximum temperature. On the other hand, 
as it can be seen in the bottom row, the polarization fraction tends to increase 
as the photon flux declines, apart from photons leaving the disc surface at 
small inclination angles, which (as already discussed) are practically unpolarized. 
Moreover, still taking as reference the plots in Figure \ref{figure:tempdens1}, 
some similarities can be observed between the behaviors of the polarization 
degree and the total density as functions of the radial distance. This further 
confirm the previous findings that ascribe the rise in polarization fraction 
to absorption effects. 
In fact, in the Schwarzschild BH case $\Pi$ increases dramatically going towards 
the ISCO, where temperature is lower, since atoms are less ionized in this region 
and the density is quite large. Then, after a minimum attained in 
correspondence to the maximum of the temperature distribution (at $\sim 10\,
\rg$), the polarization degree rises again, although the density is more or 
less constant, as the temperature decreases (and the fraction of ionized atoms 
drops). In the maximally-rotating BH case, instead, the polarization fraction 
radial profile is mostly monotonic, going from the minimum value (attained close 
the ISCO) and essentially continuing to grow up to the outer boundary. This is 
due to both the temperature decrease and the density increase, which are visible 
in the right panel of Figure \ref{figure:tempdens1}. The only exceptions occur 
very close to the ISCO, before the photon flux peak is reached, and farther 
than $\sim 20\,\rg$, with a slight decrease of the polarization fraction for 
the highest inclinations ($\ga 80^\circ$). While the former effect is analogous 
to that just discussed for small radii in the $a=0$ case, the latter can be more 
likely ascribed to the absorption features which are present at low energies 
in the behavior of $\Pi$. In fact, according to the adopted radial profile, 
the surface temperature far from the central BH turns out to be quite low ($\sim 
0.3$ keV), so that only low-energy photons are expected to contribute significantly 
at those distances. Recalling that the plots in Figure \ref{figure:IPr1} have been
obtained summing the Stokes parameters in the $2$--$8$ keV energy range, it is 
reasonable to conclude that the slight decline of $\Pi$ at large $r$ and high 
inclinations is due to the deep features which characterize the polarization 
degree at around $2$--$3$ keV (see the bottom-right panel of Figure 
\ref{figure:iPE1}), which are indeed more relevant for high values of $\theta$.

\subsection{Exploring different luminosities and optical depths}
\label{subsec:res2}
\begin{figure*}
\begin{center}
\includegraphics[width=17.5cm]{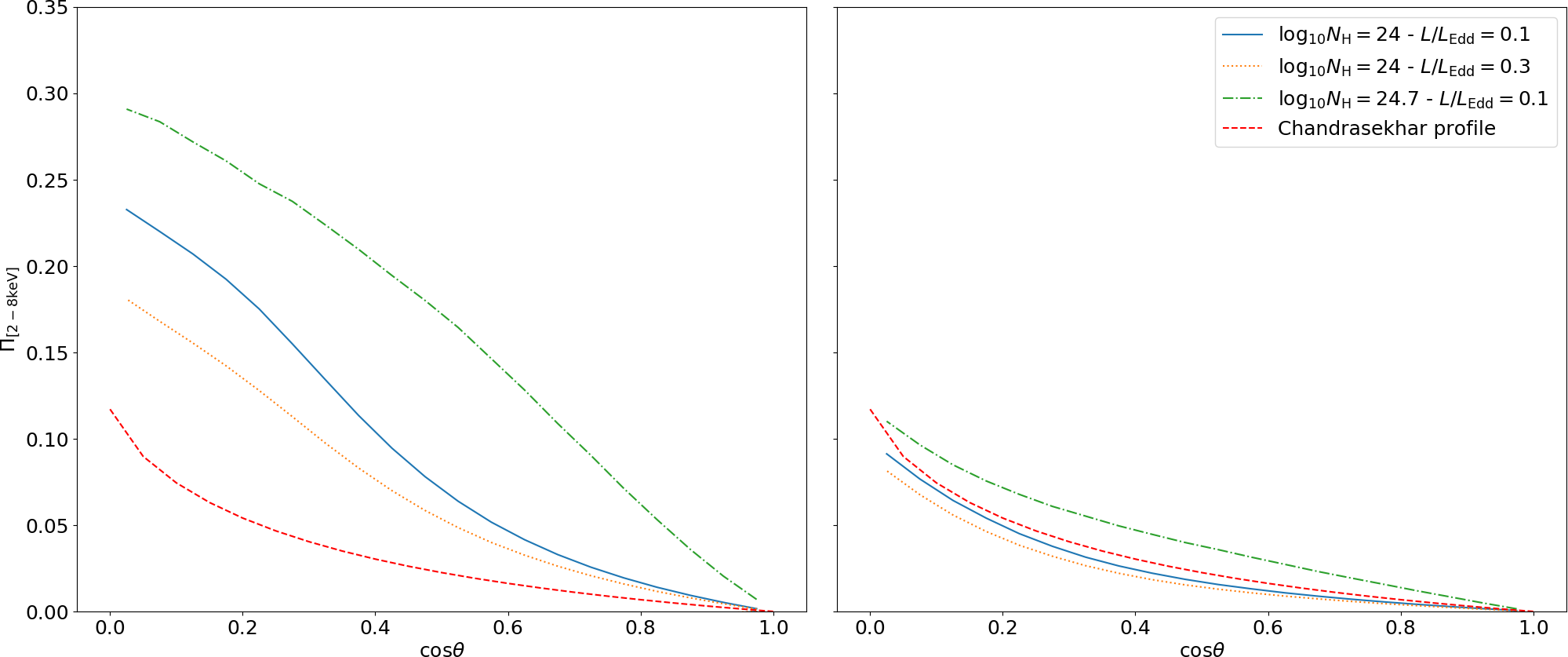}
\caption{Polarization degree plotted as a function of the cosine of the inclination 
angle $\theta$ in the two cases $a=0$ (left-hand column) and $a=0.998$ (right-hand 
column), for $N_{\rm H}^{\rm stop}=10^{24}\,{\rm cm}^{-2}$ -- $L=0.1\,L_{\rm Edd}$ 
(blue, solid), $N_{\rm H}^{\rm stop}=10^{24}\,{\rm cm}^{-2}$ -- $L=0.3\,L_{\rm Edd}$ 
(orange, dotted) and $N_{\rm H}^{\rm stop}=5\times10^{24}\,{\rm cm}^{-2}$ -- $L=0.1
\,L_{\rm Edd}$ (green, dash-dotted), while the other parameters have the same values 
as in Figure \ref{figure:iPE1}. Stokes parameters have been summed over the radial 
distance from the central BH (see equations \ref{eqn:weightedstokes}) and energy 
integrated between $2$ and $8$ keV. The red-dashed lines denotes the behavior 
calculated according the formulae by \citet[see Table XXIV]{chan60}.}
\label{figure:Pmu2}
\end{center}
\end{figure*}
After having discussed the behavior of spectral and polarization observables for
a given set of parameter values, we now test how the Stokes parameters of the 
emerging radiation can change by varying the properties of the disc material and 
the accretion flow. In this respect, we explored two further configurations, 
characterized by different values of mass accretion rate $\dot{M}$ and stop column 
density $N_{\rm H}^{\rm stop}$, leaving the BH mass unchanged at $10\,{\rm M}_\odot$ 
and the hardening factor $f_{\rm col}=1.8$. As reported in equations 
(\ref{eqn:NTtemperature}) and (\ref{eqn:n0Hinner})--(\ref{eqn:n0Houter}), varying 
the accretion luminosity acts both on the temperature and the density radial profiles. 
Figure \ref{figure:tempdens2} shows the variations of temperature and density profiles 
as a result of increasing the accretion rate from $10\%$ (i.e. the case already 
described in \S \ref{subsec:res1}) to $30\%$ of the Eddington limit, again for the 
two cases of non-rotating and maximally rotating BHs. In particular, the temperature 
undergoes a general shift upwards by a factor of $\sim 1.3$, superimposed to a radial 
displacement of the maximum towards a slightly larger distance from the center. The 
density, on the other hand, drops at all the considered radial distances by a factor 
of $\sim 3$.

The effects of such changes on the emerging flux are shown in the top row of Figure 
\ref{figure:iPE2}, where solid lines mark the behavior for $L=0.1\,L_{\rm Edd}$ (i.e. 
the case already discussed in \S \ref{subsec:res1}) and dotted lines that for $L=0.3\,
L_{\rm Edd}$ ($N_{\rm H}^{\rm stop}=10^{24}\,{\rm cm}^{-2}$ for both these runs). As 
one could expect, the photon flux attains higher values when a higher accretion luminosity 
is considered, due to the increase of the temperature all over the disc surface. For 
$a=0$ (top-left panel), the low-energy peak due to primary photons is sensibly broadened 
with respect to the $L=0.1\,L_{\rm Edd}$ case. Moreover, the spectral features ascribed 
to absorption are less pronounced, as a result of both the temperature increase (which 
also increases the ionization fraction in the disc material) and the lower values of 
density. Also in the $a=0.998$ case (top-right panel), spectra for $L=0.3\,L_{\rm Edd}$
turn out to be harder than for lower luminosities. In this case no substantial
differences can be observed in the absorption features, since, as noted before,
already at $L=0.1\,L_{\rm Edd}$ temperature was sufficiently high to significantly 
reduce absorption effects. Dash-dotted lines in Figure \ref{figure:iPE2} mark, instead,
the behavior of the emerging flux for $N_{\rm H}^{\rm stop}=5\times10^{24}\,{\rm cm}
^{-2}$, which corresponds to a Thomson optical depth $\tau \simeq3.33$. In order to 
display only the effect of changing $N_{\rm H}^{\rm stop}$, for this simulation we 
returned to $L=0.1\,L_{\rm Edd}$, so that temperature and density profiles are those 
reported in Figure \ref{figure:tempdens1}. Contrary of what happens by increasing the 
accretion luminosity, in this case the number of emerging photons turns out to be pretty 
lower in the entire energy range. The peaks of the spectral distributions fall at the 
same energy as in the original case (see \S \ref{subsec:res1}), as a result of the 
choice to adopt the same temperature profile. However, both in the Schwarzschild and 
in the maximally-rotating case absorption features are dramatically more significant, 
this effect being more evident for $a=0$. Indeed, assuming that photons escape the 
layer at a $z_{\rm max}$ corresponding to a larger optical depth implies that they 
are still involved in a conspicuous number of scatterings, so that a lower number of 
emerging photons at that altitude $z_{\rm max}$ can be reasonably expected. Moreover, 
since an increase of the optical depth translates into a decrease of the photon mean 
free path inside the disc material, this also justifies the increase of absorption 
effects, despite the fact that temperature and density are unchanged with respect to 
the initial case (with $N_{\rm H}^{\rm stop}=10^{24}\,{\rm cm}^{-2}$).

Much in the same way as the spectra, the polarization degree of the emerging
radiation is influenced as well by the changes in luminosity and optical depth. 
The bottom row of Figure \ref{figure:iPE2} shows the energy-dependence of $\Pi$, 
plotted for the same values of the parameters adopted for the flux in the top 
row. By increasing $L$ to $0.3\,L_{\rm Edd}$, the overall behavior turns out 
to be in general lowered by $\sim 1$--$2\%$ with respect to the initial case, 
with a more important reduction at lower energies and for high values of $\theta$.
On the other hand, substantial differences can be observed when the value of 
$N_{\rm H}^{\rm stop}$ is increased. In particular, a steep rise occurs as the 
inclination angle increases, attaining a value $\sim 40\%$ for $a=0$ and $\sim 
20\%$ for $a=0.998$ at $10$ keV\footnote{We notice that the values of $\Pi$ obtained 
in the simulations reported in Figure \ref{figure:iPE2} for $a=0$ can be affected 
by the poor statistics due to the low number of photons at around $2$ keV and at 
high energies.}. To explore this behavior more in depth, we reported in Figure 
\ref{figure:Pmu2} the plots of the energy-integrated polarization fraction 
as a function of the viewing inclination, similarly to those shown in Figure 
\ref{figure:Pchimu1}; also in this case the polarization degree predicted by 
Chandrasekhar's (\citeyear{chan60}) prescription is displayed for comparison. 
By increasing the accretion rate (dotted lines), $\Pi$ turns out to be substantially 
reduced at high inclinations for $a=0$ (down to $\sim75\%$ of the value attained 
for $L=0.1\,L_{\rm Edd}$), but still remaining above the level set by the Chandrasekhar's 
profile. For $a=0.998$, instead, $\Pi$ is only slightly lower than in the initial 
case, similarly to what we have noted in Figure \ref{figure:iPE2}, with an overall 
reduction not larger than $\sim1\%$. On the contrary, if the stop column density 
is increased to $5\times10^{24}\,{\rm cm}^{-2}$ (maintaining the luminosity at 
$0.1\,L_{\rm Edd}$) then the polarization degree in general increases too (dash-dotted 
lines). Also in this case the most relevant change occurs in the Schwarzschild 
limit (attaining a maximum value close to $30\%$ at high inclinations), while 
the curve is not far from the original one (and from that given by the Chandrasekhar's 
profile) in the maximally-rotating limit. All these behaviors comply with those 
just discussed for the flux. In fact, the temperature increase (with the consequent 
ionization fraction rise) and the density decrease which occur taking $L=0.3\,
L_{\rm Edd}$ both determine a lower influence of absorption effects, so that the 
polarization fraction turns out to be lower with respect to the case with a lower 
luminosity. On the other hand, considering a larger value of the stop column density 
means that photon transfer is calculated for larger optical depths. This translates 
into more significant absorption effects which produce an increase of the polarization 
degree. That being said, what we have concluded in the reference case (see \S 
\ref{subsec:res1}) holds true also for different luminosities and column densities, 
i.e. radiation is basically more polarized for $a=0$ than for $a=0.998$, confirming 
that absorption is more important for slowly rotating BHs, while scattering dominates 
as the BH spin increases.

\section{Discussion and conclusions}
\label{section:discussion}
In this paper we discussed a method to simulate the spectral and polarization
properties of radiation emitted from stellar-mass BH accretion discs in the 
soft state. Contrary to previous works, where, following the Chandrasekhar's 
(\citeyear{chan60}) prescription, polarization was considered to be due only 
to scattering of photons onto electrons in the disc, in our model we include
a self-consistent treatment of absorption effects in the disc material. To 
this aim, we solved the ionization structure in the surface layer of the disc
using the photo-ionization code {\sc cloudy} \cite[][]{fer+17}, assuming the 
Novikov \& Thorne (\citeyear{nt73}) temperature profile and the Comp\`{e}re 
\& Oliveri (\citeyear{co17}) density profile. Seed radiation (assumed to be 
unpolarized) is injected isotropically from the bottom of this layer, according 
to a blackbody distribution at the local temperature. The radiative transfer 
of photons is then calculated using the ray-tracing, Monte Carlo code {\sc 
stokes} \cite[][]{mar18,gg07}. The emerging photon Stokes parameters are 
determined for a number of virtual detectors on the surface of the layer, 
displaced at different inclinations with respect to the disc axis.

The main findings of our present investigation can be summarized in the 
following points.
\begin{itemize}
\item Placing the emitting source at the base of the surface layer forces the
polarization direction of most of the emerging photons to be parallel to the
plane of the disc (see Figure \ref{figure:scattonly}). This differs from the 
results of previous works \cite[e.g.][with emission region placed in the middle 
of the slab]{dov+08}, according to which a consistent fraction of photons is 
polarized parallel to the disc symmetry axis for $\tau\la2$. In fact, when
only scattering is considered, the polarization degree at lower $\tau$ is quite 
small for emitting source on the bottom with respect to the case of emission 
from the middle of the layer. As expected, the polarization fraction tends 
to the angular behavior described by Chandrasekhar's (\citeyear{chan60}) 
formulae at high optical depths ($\tau\ga5$).
\item We emphasize that including absorption effects goes in general towards 
increasing the polarization degree (see Figure \ref{figure:iPE1} and 
\ref{figure:Pchimu1}, left columns). This is particularly encouraging, considering 
that all the theoretical models developed so far \cite[which account for scattering 
as the only responsible for photon polarization, see e.g.][]{dov+08,sk09,tav+20} 
predict a level of polarization of $\sim 10\%$ at most. 
\item The photon spectrum and the polarization degree turn out to strongly
depend on the spin of the central BH. This can be essentially ascribed to the
fact that, as the BH spin increases, the inner boundary of the disc extends to 
smaller distances from the horizon, where temperatures are higher and more
energetic photons can be emitted (see Figures \ref{figure:tempdens1} and 
\ref{figure:IPr1}). In addition, according to the profile by \citet{co17}, the 
density turns out to be higher for lower values of $a$ (see again Figure 
\ref{figure:tempdens1}). As a consequence, photon polarization can be reasonably
expected to be higher for slowly rotating BHs, for which absorption effects are
more important, while polarization tends to approach the Chandrasekhar's profile 
for $a=0.998$ (Figure \ref{figure:Pchimu1}, right panel), where scattering effects
are dominant. 
\item Including self-consistently the absorption effects in our model allows 
us to specify the non-trivial behavior of the seed photon Stokes parameters 
as a function of photon energy, observer's inclination and radial distance
from the center, as it can be seen in Figure \ref{figure:iPE1}, \ref{figure:features}
and \ref{figure:IPr1}. 
\item We could test the dependence of spectra and polarization on the physical 
properties of the disc, such as the accretion luminosity and the hydrogen column 
density of the disc surface slab (see Figures \ref{figure:iPE2} and \ref{figure:Pmu2}). 
In particular, we have checked that increasing the accretion luminosity acts 
in increasing the photon flux and decreasing the polarization degree, as a 
result of the correspondent temperature rise and density drop (see Figure
\ref{figure:tempdens2}). On the other hand, placing the upper boundary of
the layer at a different height modifies as well the polarization properties,
increasing $\Pi$ as a higher value of $N_{\rm H}^{\rm stop}$ is considered. 
\end{itemize}

We stress again that the main goal of this paper is to explore the effects of 
absorption on the polarization properties of the thermal disc emission, so we 
give spectra and polarization observables of radiation without considering the 
effects of strong gravity on photon energy and trajectory; general relativistic 
corrections, including returning radiation, will be properly addressed in a future 
work. We also stress that our physical assumption of an ionized slab above a 
standard, black body emitting disc is certainly simplistic. A global, completely 
self-consistent treatment of the disc structure including polarized radiative 
transfer is a very ambitious goal, and clearly beyond the scope of this work. 
Nevertheless, the results of the present investigation show that including absorption 
effects alongside those of scattering is crucial for correctly modeling the 
polarization properties of emission from stellar-mass BH accretion discs, an 
important result in view of the forthcoming X-ray polarimetry missions that will 
be launched in the next decade.

\section*{Acknowledgments}
         {We thank the anonymous referee for comments which helped us improving the clarity
           of the paper.}
RT, GM and SB acknowledge financial support from the Italian Space Agency (grant 
2017-12-H.0).
MD acknowledges the support by the project RVO:67985815 and the project LTC18058.
WZ would like to thank GACR for the support from the project 18-00533S.

\section*{Data availability}
The data underlying this article will be shared on reasonable request to the corresponding 
author.

\appendix
\section{Density radial profile in the disc} \label{appendix:densities}
The equatorial density of the disc material can be expressed as follows, in
each of the three different regions in which the disc is divided, according
to which component between gas and radiation dominates in the pressure
and which process between scattering and free-free dominates the opacity
\cite[][]{co17,tav+20}:
\begin{itemize}
\item in the inner region, where radiation pressure and scattering opacity 
dominate over gas pressure and free-free opacity, respectively,
\begin{flalign} \label{eqn:n0Hinner}
& n_0({\rm H})_{\rm inn}=(1.50\times10^{19}\,{\rm cm}^{-3})\,\alpha^{-1}
\left(\frac{M}{3{\rm M}_\odot}\right) \nonumber & \\
&\,\,\,\,\,\,\times\left(\frac{\dot{M}}{10^{17}\,{\rm g\,s}^{-1}}\right)^{-2}
\xi^5\mathcal{C}^3\mathcal{D}^{-1}\mathcal{R}^2
\mathcal{P}^{-2}\,; &
\end{flalign}
\item in the middle region, where scattering is still the main source of opacity
but gas pressure dominates over radiation pressure,
\begin{flalign} \label{eqn:n0Hmiddle}
& n_0({\rm H})_{\rm mid}=(4.86\times10^{24}\,{\rm cm}^{-3})\,\alpha^{-7/10}
\left(\frac{M}{3{\rm M}_\odot}\right)^{-11/10} \nonumber & \\
&\,\,\,\,\,\,\times\left(\frac{\dot{M}}{10^{17}\,{\rm g\,s}^{-1}}\right)^{2/5}
\xi^{-37/10}\mathcal{C}^{3/10}\mathcal{D}^{-7/10}\mathcal{R}^{1/2}\mathcal{P}^{2/5}\,; &
\end{flalign}
\item in the outer region, where gas pressure and free-free opacity dominate
over radiation pressure and scattering opacity, respectively,
\begin{flalign} \label{eqn:n0Houter}
& n_0({\rm H})_{\rm out}=(3.06\times10^{25}\,{\rm cm}^{-3})\,\alpha^{-7/10}
\left(\frac{M}{3{\rm M}_\odot}\right)^{-5/4} \nonumber & \\
&\,\,\,\,\,\,\times\left(\frac{\dot{M}}{10^{17}\,{\rm g\,s}^{-1}}\right)^{11/20}
\xi^{-43/10}\mathcal{C}^{3/20}\mathcal{D}^{-7/10}\mathcal{R}^{17/40}\mathcal{P}^{11/20}\,. &
\end{flalign}
\end{itemize}
In equations (\ref{eqn:n0Hinner})--(\ref{eqn:n0Houter}) we assumed the mean rest 
mass per barion $m_{\rm b}$ as $1/56$ the mass of an $^{56}{\rm Fe}$ atomic 
nucleus \cite[][]{tb17} and $\alpha$ is the disc parameter \cite[see][we take 
$\alpha=0.2$ throughout the paper]{ss73}, while complete expressions for the 
functions $\mathcal{C}$, $\mathcal{D}$, $\mathcal{R}$ and $\mathcal{P}$ are 
reported in \citet{co17}.

\label{lastpage}

\end{document}